\documentclass[conference]{IEEEtran}
\IEEEoverridecommandlockouts
\usepackage{cite}
\usepackage{amsmath,amssymb,amsfonts}
\usepackage{graphicx}
\usepackage{textcomp}
\usepackage{xcolor}
\usepackage{algorithm}
\usepackage{algpseudocode}
\usepackage{lipsum}
\usepackage{multirow}
\usepackage{diagbox}
\usepackage{float}
\usepackage{xcolor, soul}
\sethlcolor{yellow}

\def\BibTeX{{\rm B\kern-.05em{\sc i\kern-.025em b}\kern-.08em
    T\kern-.1667em\lower.7ex\hbox{E}\kern-.125emX}}
\begin{document}

\title{
A Low-Dimension High-Fidelity Reliability Representation for 
         Reliability-based \\
         Machine Learning Modeling of \\
         Physical Unclonable Functions


\thanks{*Authors listed alphabetically with equal contribution. 

The research was supported in part by the National Science Foundation under grant No.\,\,2103563.}
}

\author{\IEEEauthorblockN{Gaoxiang Li* and Yu Zhuang*}
\IEEEauthorblockA{Department of Computer Science \\
Texas Tech University\\
Lubbock, TX 79409, USA \\
gaoli@ttu.edu, yu.zhuang@ttu.edu}
}

\maketitle

\begin{abstract}
Physical Unclonable Functions (PUFs) are emerging as promising security primitives for IoT devices, providing device fingerprints based on physical characteristics. Despite their strengths, PUFs are vulnerable to machine learning (ML) attacks, including conventional and reliability-based attacks. Conventional ML attacks have been effective in revealing vulnerabilities of many PUFs, and reliability-based ML attacks are more powerful tools that have detected vulnerabilities of some PUFs that are resistant to conventional ML attacks. 
Reliability-based ML attacks leverage information of PUFs' unreliability, and due to the existence of many reliability-enhancing techniques, we were tempted to examine the feasibility of improving resistance against reliability-based ML attacks using some reliability enhancing techniques. Majority voting is a simplistic technique and was found experimentally  that it substantially improve PUFs' ability to withstand, though not proven to defeat, those attacks. 
It is known that majority voting reduces but does not eliminate unreliability, we are motivated to investigate if new attack methods exist that can capture the low unreliability of PUFs, which led to the development of a new reliability representation and the new representation-enabled attack method that has experimentally cracked PUFs enhanced with majority voting of high repetitions.
\end{abstract}


\section{Introduction}

Physically Unclonable Functions (PUFs) have emerged as a promising security primitive for low-cost  low-power Internet-of-Things (IoT) devices.  By providing devices with unique fingerprints derived from their inherent physical characteristics, PUFs offer a new approach to implementing authentication and encryption. The foundational concept behind PUFs is to move away from the conventional practice of storing keys in non-volatile memories. Not only because such memories are costly, but they are also vulnerable to specific attacks, such as cold boot attacks \cite{halderman2009lest}, especially when the IoT devices are easily accessible by the crowd. 

Instead of retrieving keys from non-volatile memories, PUFs utilize inherent physical variations as devices'  ``fingerprints''. These fingerprints, arising from minute inconsistencies during manufacturing processes, are physically unclonable and distinct for each device. This inherent uniqueness is utilized to generate digital identifiers, which can serve as a robust defense mechanism against potential security threats, including impersonation and counterfeiting, prevalent in IoT devices \cite{guajardo2007fpga,ruhrmair2009foundations,maes2010physically}. PUFs have been extensively studied in recent years \cite{lee2004technique, lim2004extracting,gassend2002controlled,gassend2002silicon,tuyls2007security}. Due to their resilience against physical clones, no requirement of secure memory for key storage, and their cost-effective hardware implementation, PUFs hold the potential to overcome challenges that conventional cryptographic methods face in the IoT landscape. 

Despite their resilience against physical cloning, PUFs are not immune to vulnerabilities, and a thorough understanding of all potential weaknesses is crucial. 
PUFs face a spectrum of threats, including side-channel-based modeling attacks and CRP-based based modeling attacks \cite{ruhrmair2013puf, aseeri2018machine, santikellur2019deep, majzoobi2008lightweight, alkatheiri2017towards, wisiol2022neural, mursi2020fast, thapaliya2021machine}, where CRP denotes challenge-response-pair. 

CRP-based modeling attacks include conventional and reliability-based machine learning (ML) attacks. While some PUFs with sophisticated circuit architectures can resist conventional ML attacks, reliability-based ML attacks have been successful in compromising many PUFs that conventional methods cannot breach \cite{becker2015gap,6581579}.  Reliability-based attacks exploit the variability in PUF responses, by applying the same challenges repeatedly to the PUF to get multiple responses for each challenge and analyzing the responses to uncover exploitable patterns.  Since PUFs inherently possess some level of unreliability, the risk of susceptibility to reliability-based attacks is real, even for PUFs engineered to withstand conventional machine learning assaults. Resistance to reliability-based modeling attacks is hence an important metric for evaluating PUF designs.

Some PUF attack methods use side-channel information, including power consumption \cite{mahmoud2013combined,ruhrmair2013power,wei2014reverse, becker2014active}, electromagnetic waves \cite{merli2013localized}, photonic emission \cite{tajik2017photonic}, signal timing or frequency \cite{wei2014reverse,xu2014hybrid}, etc. to help predict PUFs' responses. There have been many successful efforts on fighting side-channel attacks, including dynamic voltage scaling \cite{yang2005power,avirneni2013countering}, balanced gates \cite{mahmoud2013combined,yu2019leveraging}, converter reshuffling \cite{yu2015time,yu2015charge}. Some side-channel attack methods use reliability information together with other side-channel information \cite{liu2022multiclass,gao2023mlmsa,9973338}. In this paper we consider attacks using only reliability information.

Since there are many techniques \cite{gu2017improved,sahoo2017multiplexer,xu2016clockless,uddin2016techniques,patil2017improving,yamamoto2015new,xu2016using,wen2017enhancing,amsaad2021efficient,xu2023modeling,he2020highly, lu2021high,lin2022enhancing,8438897} developed for improving the reliability of PUFs, it is natural to think if reliability-enhancing techniques can also be used to fight,or at least to reduce the effectiveness of, reliability-based ML attacks.



Majority-voting is a simplistic reliability-enhancing technique, and we decide to test it on PUFs that have been shown to have succumbed to reliability-based ML attacks. We implemented PUFs on FPGAs and applied majority-voting to PUF responses with different repeats of majority-voting. Then, reliability-based ML attacks were applied to responses out of majority-voting enhanced PUFs. In the study, we have discovered that attack methods exhibit decreasing attack power as PUF reliability increases and 
become rarely successful when majority voting uses 50 or more repeats with the number of CRPs used in experimental attack studies.
\par

To understand how majority-voting reduces the power of reliability-based attack methods,
we started to examine existing reliability-based ML attack methods, and some results led us to believe that, for a group of broadly applicable methods, the representation of reliability information is the underlying factor for the failure of these methods for attacking high reliability PUFs. With their reliability representation, the dimensionality of the output of a machine learning method grows with the reliability of the PUF, and eventually destroys the learning power of the method when the reliability, and consequently the dimensionality, becomes too high. In an attempt to develop new reliability-based ML attack methods for highly reliable PUFs, we introduce a novel approach to representing PUF reliability, an approach towards lower output dimensionality for machine learning methods with no or limited decrease in reliability information. With the new representation of reliability, neural network methods exhibit substantially improved learning power for attacking highly reliable PUFs. Thus, the main contributions of this paper are 
\begin{enumerate}
    \item the discovery that majority voting 
    reduce the effectiveness of 
    existing reliability-based ML attack methods, and 
    \item the development of a new representation of PUF reliability information, which enables neural network methods to successfully attack highly reliable PUFs.
\end{enumerate}

The remainder of the paper is organized as follows. Section~\ref{sec3} presents a brief survey of existing reliability-based machine learning attack methods, including an experimental comparative study of three surveyed attack methods. Section~\ref{sec4_defense} reports our study on defending existing reliability-base ML attacks using majority voting. In Section~\ref{sec_LDHF}, we explore reasons for the effectiveness of majority voting in fighting a group of existing reliability-based ML attack methods, and propose a new representation for experimentally measured reliability for neural network attack methods. Section~\ref{sec_experiment} presents experimental study of attack methods equipped with the proposed reliability representation, and Section~\ref{sec_conclusion} gives the conclusion.

\section{Background on Arbiter PUF and its variants}
In order to help technical discussions in later sections, we will briefly describe the Arbiter-PUF, XOR-Arbiter-PUF and Interpose-PUF in this subsection.

\subsection{Arbiter PUF}
\begin{figure}
        \includegraphics[width=9.5cm]{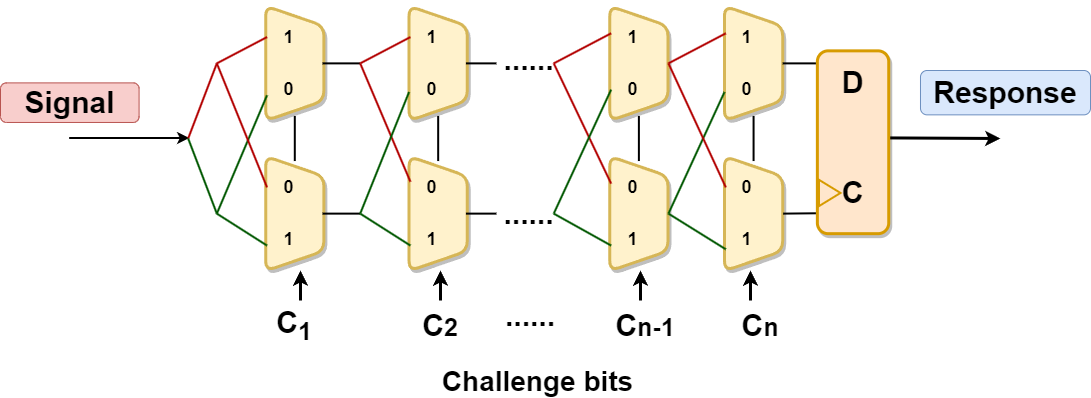}
        \caption{An arbiter PUF with n bits of challenge.}
        \label{f1}
\end{figure}

An n-bit arbiter PUF, shown in Figure \ref{f1}, consists of n stages, each containing two multiplexers (MUXs). When a rising signal is applied, it enters the arbiter PUF at the first stage and splits into two signals. These two signals are then routed through gates at each stage, with the propagation paths at each stage determined by the challenge bit. Finally, the two signals reach a D flip-flop, which serves as an arbiter to determine whether the signal on the top path or the signal on the bottom path arrives first. If the top path signal arrives first, the D flip-flop returns a response of 1, and if the bottom path signal arrives first, the D flip-flop returns a response of 0.

\subsection{XOR-Arbiter-PUF}
\begin{figure}
       \centering
        \includegraphics[width=8cm]{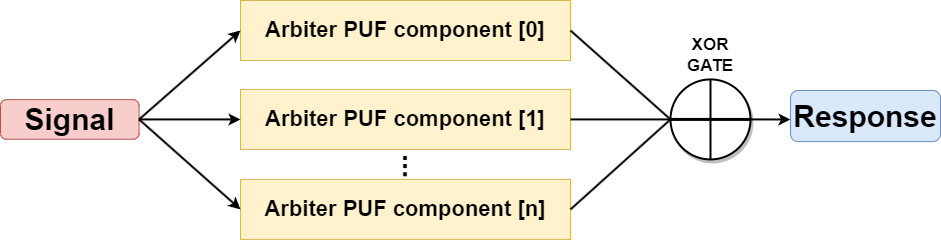}
        \caption{Illustration of an XOR-PUF with n arbiter PUF components; the final response is the XORed result of n arbiter-PUF responses.}
        \label{f2}
\end{figure}

The XOR-arbiter-PUF (XOR-PUF) In \cite{suh2007physical} was proposed to overcome the susceptibility of arbiter PUFs to machine learning attacks. The \emph{k}-XOR-PUF utilizes a \emph{k}-input XOR gate to combine the responses of $k$ arbiter PUFs to generate the final response. An XOR-PUF example is shown in Figure \ref{f2}.

\subsection{Interpose-PUF}

 \begin{figure}
        \includegraphics[width=9.5cm]{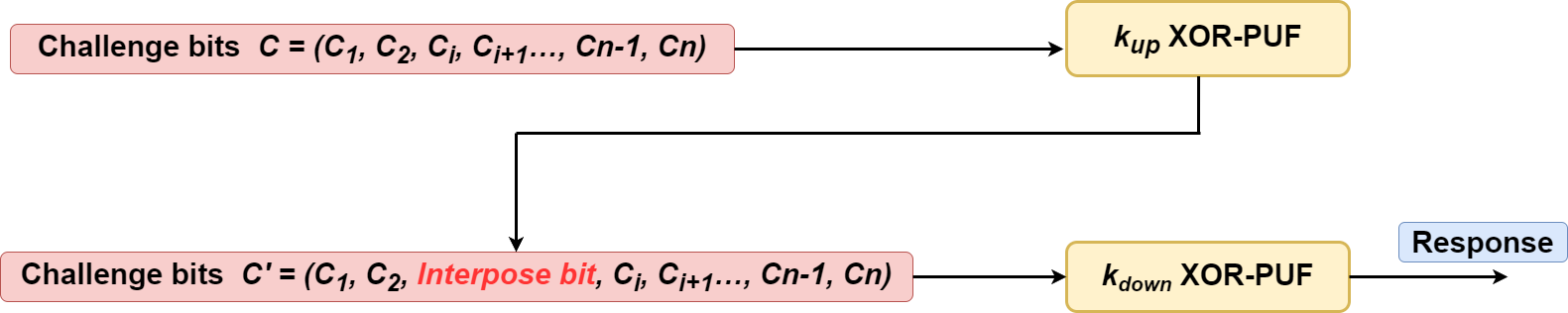}
        \caption{Schematic representation of Interpose PUFs.}
        \label{f3}
\end{figure}

Nguyen et al. \cite{nguyen2019interpose} introduced the Interpose-PUF (IPUF) that utilizes a domino structure to resist reliability-based machine-learning attacks without the need for countermeasures at the protocol level. The IPUF comprises two XOR Arbiter PUFs, namely the x-(upper) and y-(lower) PUF. The x-XOR PUF response serves as an additional challenge bit for the y-XOR PUF. The authors argue that the constituent APUFs of the x-XOR PUF cannot be accurately modeled by reliability attacks, based on their security analysis.

\section{Brief Survey of Reliability-based Machine Learning Attacks} \label{sec3}
 \begin{figure*}
       \centering
        \includegraphics[width=12cm]{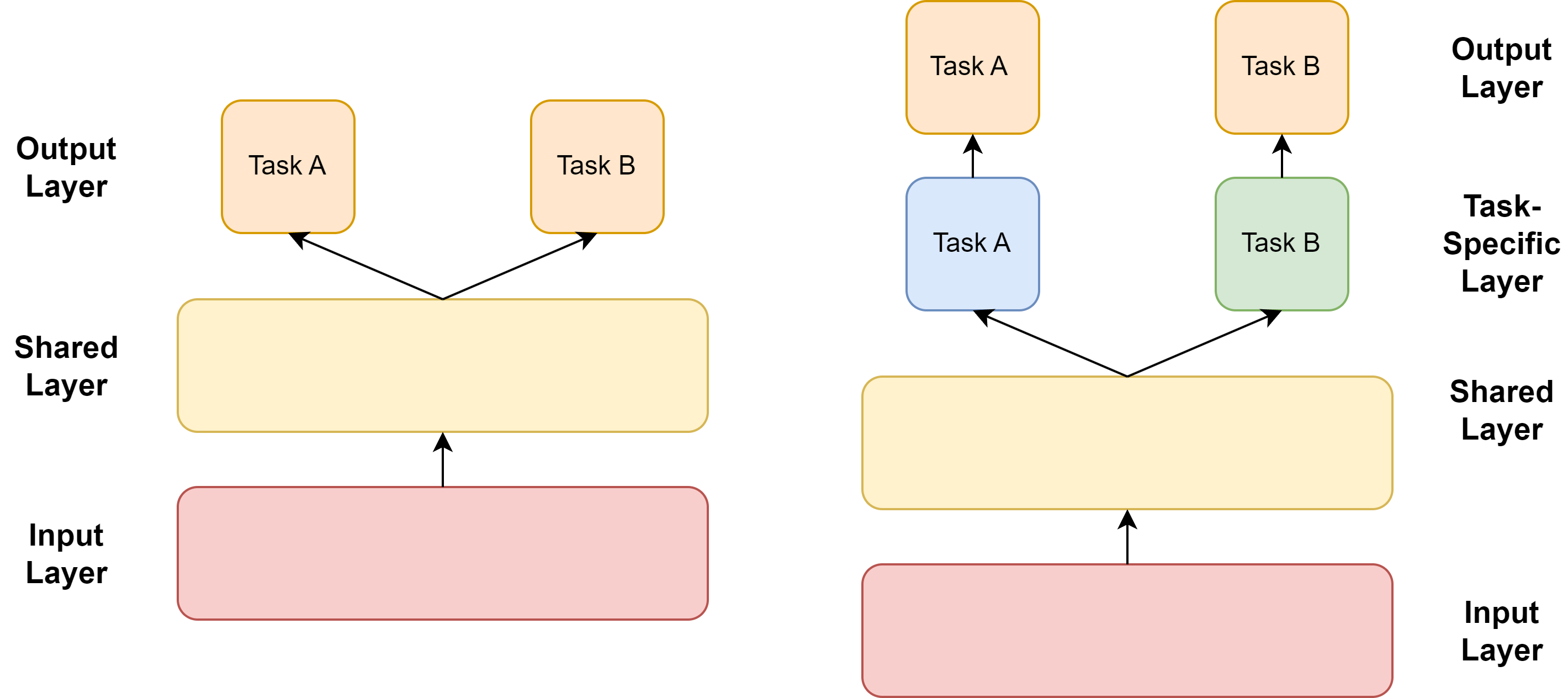}
        \caption{Illustration of the neural network architectures of the MLMSA \cite{gao2023mlmsa}  (left) and the ALScA \cite{9973338} (right)}
        \label{f6}
\end{figure*}

\subsection{PUF Circuit Architecture-Tailored Attack Methods} 


The first attack method that utilizes PUF reliability information is the CMA-ES  \cite{becker2015gap} for attacking XOR PUFs, which models the weights of different APUF components of an XOR PUF using an evolutionary strategy. Due to the non-smoothness of the objective function with respect to the model parameters, CMA-ES does not employ gradient-based optimization and is hence slow.  Tobisch et al. \cite{tobisch2020combining} proposed a multi-objective optimization (MOO) method for attacking XOR PUFs and IPUFs by modeling every component APUF using the reliability information with a differentiable function and imposing a constraint to steer convergence to un-explored component APUFs. 
These two methods are tailor designed for the circuit structures of the targeted PUFs, 
which limit their applicability to other PUFs.

\subsection{Generally Applicable Reliability-based Attack Methods}\label{ExistingGeneralMethods}

Neural Networks (NN) have been found to be highly effective and generally applicable for attacking PUFs \cite{alkatheiri2017towards,aseeri2018machine,aseeri2018subspace,santikellur2019deep,mursi2020fast,wisiol2021neural,thapaliya2021machine}, and were also used in attacks that leverage PUF reliability information as discussed in the following.

\subsubsection{A Multiclass-based Attack Method}

The multiclass side-channel attack (MSA) method \cite{liu2022multiclass}, 
employs a neural network with the output constructed through ``feature-crossing", where the number of output bits is determined by the product of the possible numbers of values of each output feature category, including the PUF response, a reliability measurement, and a power consumption measurement. The method also works when the power consumption information is dropped, though in \cite{liu2022multiclass} no experimental study was carried out using only response and reliability information.  When it is restricted to use only response and reliability information, the MSA method \cite{liu2022multiclass} has the following major characteristics:
\begin{itemize}
    \item The reliability measurement $r_m$ used in \cite{liu2022multiclass} can choose the number of 1's when a challenge is repeatedly applied $m$ times, so the reliability measurement $r_m$ has $(m\!+\!1)$ possible values. Since the PUF response $r$ has two possible values (0 or 1), the neural network has $2(m\!+\!1)$ output bits representing $2(m\!+\!1)$ possible outcomes for the random variable $(r, r_m)$ which is obtained after applying the challenge once to obtain $r$ and $m$ times to obtain $r_m$. 
    \item The input to the neural network is a challenge, and the output employs one-hot encoding where the $2(m\!+\!1)$ output bits assume the value of 0 or 1 with only one output bit being 1, and the 1-valued output bit indicates that the random variable $(r, r_m)$ assumes one of the $2(m\!+\!1)$ possible outcomes.
\end{itemize}

\subsubsection{The Multi-label Multi-side-channel-based Method}

The Multi-label Multi-side-channel-based (MLMSA) attack method \cite{gao2023mlmsa} is another neural network method with different groups of output bits representing different types of information, a 1-bit group for the PUF response and a group for each different type of side-channel information. Like the MSA \cite{liu2022multiclass}, the MLMSA method uses response, a reliability measurement, and power consumption measurement, and also works when only response and a reliability measurement are used. When it is restricted to use only response and reliability information, the output of the MLMSA neural network method \cite{gao2023mlmsa} contains two types of output bits:
\begin{itemize}
    \item one response bit and $(m\!+\!1)$ reliability-indication bits for attacks where each challenge is repeatedly applied $m$ times, where
    \item the response bit $r$ takes the response of the PUF when the challenge is applied once to the PUF,
    \item only one of $(m\!+\!1)$ reliability-indication bits is 1, where the position of the 1-valued bit indicates the number of times the $m$ PUF responses being 1, and
    \item all output nodes, including the node for the response bit and the nodes for reliability-indication bits, take as input the output from the same neural network layer that is immediately preceding the output layer, but uses different activation functions for the response bit and the reliability-indication bits.
\end{itemize}
The last of the four properties listed above for the MLMSA method means that the two groups of output bits share the same neural network body from the input layer to the layer that precedes the output layer (see left of Fig.~\ref{f6} for illustration when PUF response and reliability are treated as two tasks with Task A representing response and Task B representing reliability information).

Experimental attack data reported in \cite{gao2023mlmsa} show that the MLMSA method \cite{gao2023mlmsa} has similar performance with the  \cite{liu2022multiclass} for tested PUFs of medium circuit architecture sizes, but exhibits higher learning power than the MSA method \cite{liu2022multiclass}, when applying to PUFs of large PUF architecture sizes (e.g. XOR PUFs with large XOR gate size and IPUF with large number of component PUFs).

\subsubsection{The Auxiliary Learning Side-channel Attack Method}

Like the MLMSA, the Auxiliary Learning Side-channel Attack (ALScA) method \cite{9973338} uses response, a reliability measurement, and a power measurement. But the ALScA has different sub-networks for different groups of output bits, as illustrated in Fig.~\ref{f6} that compares the neural network body structures of the MLMSA and the ALScA when using only the PUF response and reliability information. 

It is our belief that response and reliability represent different aspects of the PUF behavior and may need larger sub-networks than only one output layer (e.g. the left network architecture in Fig.~\ref{f6}) in order to accurately capture the two aspects of PUF behavior, prompting us to surmise that the ALScA may have superior modeling performance than the MLMSA.  The study \cite{9973338} did not conduct comparative study of ALScA and MLMSA, but presented data for comparisons between ALScA \cite{9973338} and the MSA \cite{liu2022multiclass} using power information. 
Hence, it is not clear how ALScA compares with the MLMSA and other reliability-based attack methods when only PUF response and reliability information are used.

\subsection{Experimental Comparative Study of Existing Methods}\label{sec3.C}

We believe further attack studies are needed to examine the performances of the methods surveyed in earlier subsections 
when only response and reliability information are available. Thus, we are going to perform experimental attack studies to compare the attacking powers of the MSA, MLMSA and the ALScA methods
and the PUF circuit-tailored CMA-ES method and multi-objective optimization (MOO) method of Tobisch et al.\cite{tobisch2020combining}. 

\subsubsection{Experimental Tools and Setup}

To evaluate the selected attack methods, we used the Pypuf simulator \cite{wisiol2021neural} to create a set of PUF instances. 
For each PUF type examined, we generated 20 unique simulated instances with different seeds to maintain reproducibility. Additionally, a Gaussian-distributed term $\emph{noise} \sim N (0, \sigma^2)$ is added to simulate the unreliability of PUFs in practice. The \emph{noise level} in the later section denotes the value of $\sigma^2$ added in the CRP generation (e.g. 0.05 noise in tables means $\sigma^2=0.05$ as compared with the standard deviation of the PUF delay differences which is 1).

In our experimental attack study, we split the data into 90\% for training and 10\% for testing, with 1\% of the training data used for validation. The experiments were implemented in Python 3.7, utilizing TensorFlow and Keras for machine learning tasks. We conducted the experiments on a high-performance computing cluster with 128 AMD EPYC™ 7702 cores and 512 GB of memory to ensure efficient processing.

In an experimental attack of a PUF design, starting from a small number of CRPs, we gradually increase the number of CRPs in attacks until achieving a 90\% success rate across all PUF instances or all attacks remain failed when reaching 10 million CRPs. An attack was considered successful if it achieved a testing accuracy above 85\%. Due to the 48-hour runtime limit on the computing cluster, any attack that did not converge within this time limit is considered a failed attack. This approach helped identify the approximate minimal number of CRPs for attacking a PUF.

\subsubsection{The Experimental Attack Study}

The MCA, the MLMSA, and the ALScA are neural network methods, originally developed with response, reliability, and power information. To examine their performances for reliability-based attacks, first we set to find appropriate values of their neural network architectural parameters that give good balanced performances for a set of tested PUFs. The PUFs we chose to estimate the NN architectural parameters are the 10-XOR-PUFs, the (1,8)-IPUFs, and the (6,6)-IPUFs. All PUF instances are of 64 stages, and a noise value of 0.05 was applied to simulate  PUF unreliability. For each PUF type, 20 instances were generated.

In the process, we use the PUF simulator Pypuf \cite{wisiol2021neural} to generate PUF instances by varying the number of shared and task-specific layers of the neural network architecture of the ALScA. Note that the MLMSA method is a special case of the ALScA with 0 task-specific layer for all tasks, our experimental attack-based process to look for optimizal NN architecture runs through a range of architectural parameters for the ALScA method by including cases with 0 task-specific layer.
The process revealed that the ALScA attains optimal performance with 3 shared layers and 2 task-specific layers and the MSA and MLMSA methods have the best performance with 3 shared layers. These parameters provide a balanced trade-off between the number of required CRPs, success rate, and training time, with the best balanced efficiency and effectiveness for the PUF attacks we carried out. 


With the NN architectural parameters for the NN-based methods chosen, we conducted reliability-based machine learning attacks using the ALScA method, the MOO method of Tobisch et al.\cite{tobisch2020combining}, the CMA-ES method, the MSA method, and the MLMSA method,  with only reliability information and PUF responses for the ALScA, the MSA and the MLMSA. Due to the unavailability of the MSA and MLMSA's source code, we re-implemented this method in python. We managed to replicate Tobisch's reported results for some cases and for cases we couldn't replicate Tobisch's reported results, we relied on the original data reported in the paper \cite{tobisch2020combining} of Tobisch et al. The codes developed for the experimental study, including the codes implementing the existing attacking method, will be made available for reproducibility study by peer researchers after the paper is accepted for publication.

\begin{table}[]
    \centering
    \caption{Reliability-based attacks on simulated 64-stage XOR-PUFs. The noise level was set to 0.05. In Column ``Mthd'', A denotes the ALScA, B denotes the MOO, C denotes the MLMSA, D denotes the MSA, and E denotes the CMA-ES.  }
    \linespread{1.2}\selectfont
    \setlength\tabcolsep{5pt}
    \label{simu_xpuf1}
    \begin{tabular}{|c|c|c|c|c|}
        \hline
        \textbf{k-XOR-PUF} & \textbf{Mthd} & \textbf{Traning CRPs} & \textbf{Avg. Acc.} & \textbf{Training Time} \\ \hline
        \multirow{3}{*}{6} & A & 30k & 93\% & 1 min \\ 
                           & B & 40k & 88\% & 6 min \\
                           & C & 70k & 95\% & 1 min \\ \hline
        \multirow{3}{*}{8} & A & 100k & 94\% & 2 min \\
                           & B & 60k & 88\% & 10 min \\
                           & C & 130k & 95\% & 2 min \\ \hline
        \multirow{3}{*}{10} & A & 300k & 95\% & 6 min \\
                            & B & 200k & 89\% & 1 hr \\
                            & C & 500k & 95\% & 15 min \\ \hline
        \multirow{3}{*}{11} & A & 400k & 95\% & 15 min \\
                            & B & 500k & 89\% & 40 min \\
                            & C & 1m & 95\% & 25 min \\ \hline
        \multirow{3}{*}{12} & A & 500k & 95\% & 25 min \\
                            & B & 1m & 89\% & 2.5 hrs \\
                            & C & 5m & 95\% & 1 hr \\ \hline
        \multirow{3}{*}{16} & A & 1.5m & 95\% & 50 min \\
                            & B & 5m & 89\% & 5 hrs \\
                            & C & 10m & Failed & N/A \\ \hline
    \end{tabular}
    \end{table}

\begin{table}[htbp]
    \centering
    \caption{Reliability-based attacks on 64-stage simulated IPUFs with noise level set to 0.05. 
    Note that we were only able to reproduce MOO's claimed results partially. For the results that we could not reproduce (star notation * in the table), we used the originally claimed results in the table.}
    \linespread{1.2}\selectfont
    \setlength\tabcolsep{5pt}
    \label{simu_ipuf1}
  \begin{tabular}{|c|c|c|c|c|}
        \hline
        \textbf{(x,y)-IPUF}  & \textbf{Mthd} & \textbf{Training CRPs} & \textbf{Avg. Acc.} & \textbf{Training Time} \\ \hline
        \multirow{3}{*}{1,6} & A &    180k   & 95\% & 7 min \\
                             & B &    150k   & 90\% & 15 min \\
                             & C &    200k   & 95\% & 5 min \\ \hline
        \multirow{3}{*}{1,8} & A &    360k   & 95\% & 10 min \\
                             & B &    200k   & 88\% & 2 hrs \\
                             & C &    700k   & 95\% & 20 min \\ \hline
        \multirow{3}{*}{1,10}& A &    800k   & 95\% & 35 min \\
                             & B &    800k   & 89\% & 2 hrs \\
                             & C &     2m    & 95\% & 1 hr \\ \hline
        \multirow{3}{*}{6,6} & A &    600k   & 95\% & 17 min \\
                             & B &    300k*  & 84\%* & 3 hrs* \\
                             & C &    1.5m   & 95\% & 40 min \\ \hline
        \multirow{3}{*}{7,7} & A &    900k   & 95\% & 45 min \\
                             & B &    600k*  & 81\%* & 7 hrs* \\
                             & C &     10m   & Failed & N/A \\ \hline
        \multirow{3}{*}{8,8} & A &     3m    &  95\%  & 6 hrs \\
                             & B &  No Data* & No Data*& No Data* \\
                             & C &     10m   & Failed & N/A \\ \hline
    \end{tabular}
    \end{table}


Data in Table~\ref{simu_xpuf1} and Table~\ref{simu_ipuf1} show that the ALScA has substantially higher performance in training time than the two other methods in all tested cases, has similar or higher performances in CRPs and Accuracy when compared with the other two,  but exhibits substantially higher performance in CRPs for large PUFs.   Given its potential broad applicability and robust and high performance, we deem the ALScA method as the most powerful among existing reliability-based machine learning attack methods.

\section{A Simplistic Defense Against Existing Reliability-Based Attacks}\label{sec4_defense}

Since many techniques \cite{gu2017improved,sahoo2017multiplexer,xu2016clockless,uddin2016techniques,patil2017improving,yamamoto2015new,xu2016using,wen2017enhancing,amsaad2021efficient,xu2023modeling,he2020highly, lu2021high,lin2022enhancing,8438897} have been developed for improving the reliability of PUFs, it is natural to think if reliability-enhancing techniques can be used to fight reliability-based ML attacks.
Majority-voting is a well-known simplistic reliability-enhancing technique, and we decided to test it on PUFs that have succumbed to existing reliability-based ML attacks. 

\subsection{FPGA Implementations of Majority-Vote-Enhanced PUFS.} 




For our study on the feasibility of building defense using majority voting, we choose to use hardware-generated PUF CRP datasets instead of simulated data for our experiments. 
Simulated data, while beneficial for fast comparison study due to their reproducibility and ease of control, lacks the complexity and unpredictability of real-world PUF behaviors. Enhanced reliability features, crucial for practical applications, are best demonstrated and tested on actual hardware to ensure the findings are robust and applicable in real scenarios. This approach underscores our commitment to validating our method's effectiveness in genuine use cases.


Our experimental PUFs were deployed on Xilinx Artix®-7 FPGAs, incorporating a configurable MicroBlaze CPU. We utilized VHDL for PUF construction and Xilinx Vivado 15.4 HL design edition for development, with PUF placement managed horizontally via Tool Command Language (TCL). Random challenges were fed to the PUFs using the Xilinx SDK, with responses recorded for each challenge.

CRP generation hinged on AXI General Purpose Input/Output (GPIO) interfaces, with one GPIO dedicated to initiating signals, another to capturing responses, and additional GPIOs for challenge delivery. Challenges were derived using a Pseudo-Random Number Generator (PRNG), defined by:
    \begin{equation}
    \label{equ:PRNG}
    C_{n+1} = (a \times C_n + g) \mod m,
    \end{equation}
where $C$ is the sequence of the generated random number, $a$ is a multiplier, $g$ is a given constant, and $m$ is $2^{K}$ where K is the number of stages. To speed up the data transfer between PUFs and the computer, AXI Universal Asynchronous Receiver Transmitter (UART) was used with a baud rate at $230,400$ bits/second. Finally, the Tera Term, which is a terminal emulator program, is used for printing and saving the device output. 
    
Furthermore, the voltage is set to 2.0 W, and the junction temperature reported by the Xilinx Artix\textregistered-7 FPGA is  $26.0^\circ C $ and the Thermal margin is ${59.0^\circ C}$(12.3W).

A majority voting scheme is incorporated as an integral part of our methodology in our FPGA-based PUF implementations on Xilinx Artix®-7. The majority voting mechanism operates by repeatedly applying the same challenge to the PUF and then aggregating the responses to determine the most common (majority) response, which is then designated as the final output for that particular challenge. Specifically, for each challenge applied to the PUF, we executed multiple iterations of challenge-response cycles. The majority vote was computed directly on the FPGA using custom VHDL logic, designed to count and compare the frequency of each response received for a given challenge, ultimately selecting the response with the highest frequency as the final output.

\begin{figure}
       \centering
        \includegraphics[width=8cm]{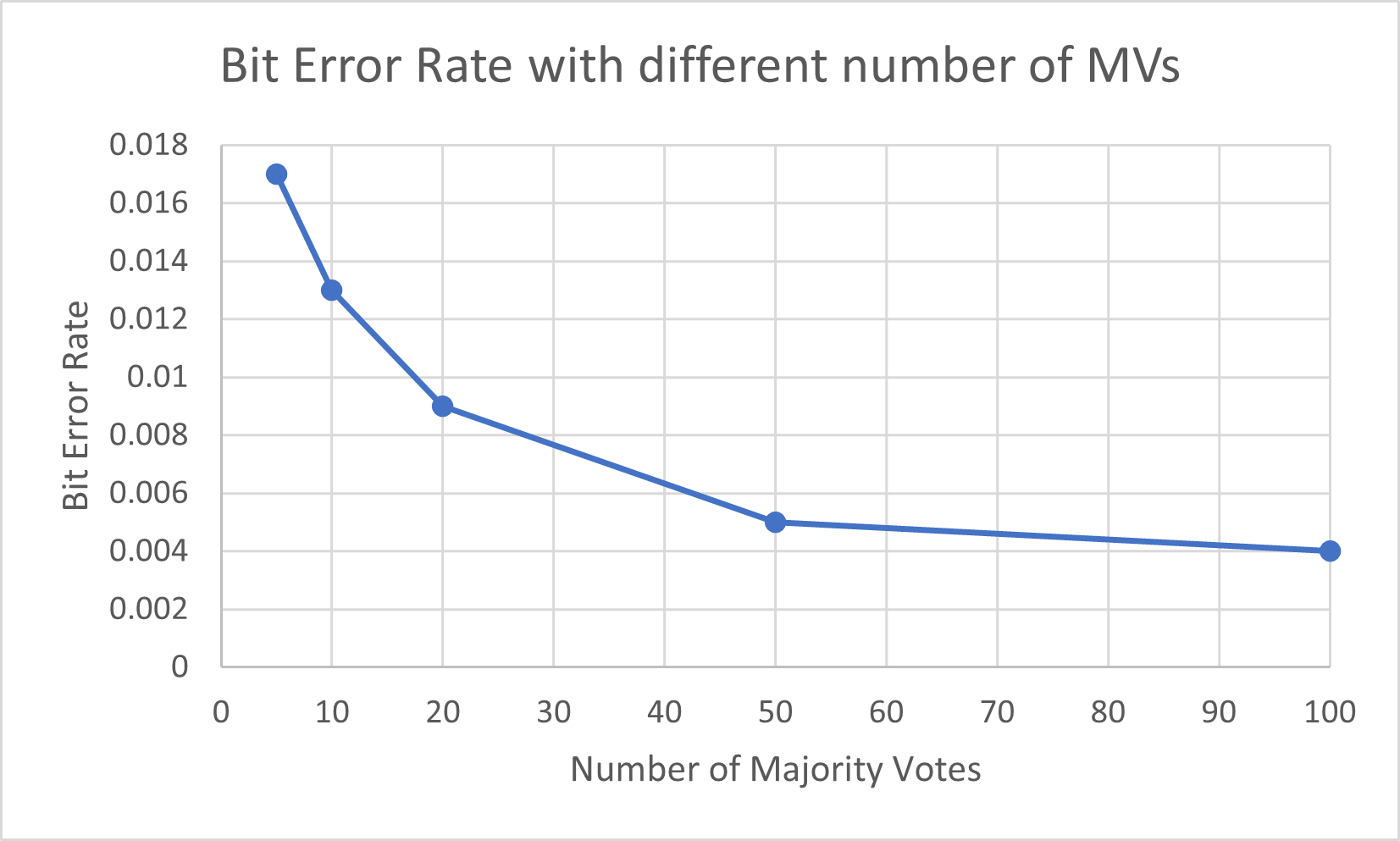}
        \caption{PUF reliability performance under different numbers of majority votes applied}
        \label{ber}
\end{figure}

Our experiments explored various counts for majority voting, including 5, 10, 20, 50, and 100 repeats. We do not go beyond 100 since low-power devices usually have limited power budget and a majority-voting with 100+ repeats may incur significant operational overhead for such devices. 
The selection of majority vote counts (Num\_MV) at 5, 20, and 50 give rise to a range of reliability, and we evaluate the resulting reliability using Bit Error Rates (BER). 
The measured BER values for the the Num\_MV values were plotted in Fig.~\ref{ber}, clearly indicative of increasing reliability (or decreasing BER) as the number of MV repeats increases. 

Considering practical constraints, we decide to cap majority voting at 50 repetitions in our study of reliability-enhanced PUFs. This decision was informed by the diminishing returns in reliability improvement beyond this point as shown in Fig.~\ref{ber} and pragmatic considerations in real-world PUF applications, where excessive majority voting could compromise system performance and resource efficiency.

\subsection{Attack Methods Used in Experimental Studies}




In our exploration of the robustness of majority-voting enhanced PUFs against reliability-based ML attacks, we have selected two  methods for our experimental studies --- the ALScA method and Multi-objectives-optimization (MOO) method of Tobisch et al.\cite{tobisch2020combining}.  
The ALScA method represents the most recent advancement in reliability-based neural network methods for attacking PUFs.
The MOO method is the best performing PUF circuit-tailored attack method. 
Also, both these methods have been empirically validated in literature to outperform other existing techniques in their respective categories. These prior validations underscore their relevance and make them appropriate candidates for our investigation into the effectiveness of current reliability-based attack strategies, particularly against high-reliability PUFs.
    
For the ALScA method, we utilize the following parameters in our experimental setup:
\begin{itemize}
    \item \textbf{NN Architecture}: In alignment with our earlier findings, the ALScA method utilizes an optimal architecture comprising 3 shared layers followed by 2 task-specific layers. The shared layers are configured with three hidden layers containing 64, 128, and 128 neurons, respectively, ensuring a comprehensive feature extraction from the input data. The task-specific layers, each consisting of 64 neurons, are tailored to fine-tune the learning process for the distinct tasks of reliability assessment and PUF response prediction.
    
    \item \textbf{Learning Rate}: We employ an initial learning rate of 0.001 for the Adam optimizer. 

    \item \textbf{Batch Size}: A batch size of 1000 samples is selected for training, optimizing the trade-off between computational efficiency and the accuracy of gradient estimation. 

    \item \textbf{Epochs}: Training is conducted for up to 150 epochs, incorporating early stopping based on validation loss to mitigate the risk of overfitting. 

    \item \textbf{Loss Weight}: The loss functions corresponding to reliability information and PUF response are assigned weights of 1.8 and 1, respectively. This weighting scheme, mirroring that of the MLMSA method, is designed to appropriately balance the importance of each task in the overall learning objective, emphasizing the critical role of reliability information in the attack process.

\end{itemize}

For the implementation of the Multi-objectives-optimization method of Tobisch et al.\cite{tobisch2020combining} in our experimental study, we adhere strictly to the original parameter settings as specified in the publicly available code released by Tobisch. This adherence ensures the fidelity of our adaptation to the unique context of hardware-generated CRP datasets, enabling a direct and fair comparison of the method's efficacy in the realm of hardware PUF analysis.

\subsection{Experimental Set-up and Test Data, and Discussion of Test data}\label{sec_defenseTestData}

\begin{table}
\centering
\caption{Attacking majority-voting-enhanced 64-bit 10-XOR-PUFs using
 the ALScA method \cite{9973338} and the MOO method \cite{tobisch2020combining}}
\label{puf1}
\linespread{1.3}\selectfont
\setlength\tabcolsep{5pt}
\begin{tabular}{|c|c|c|c|}
\hline
\textbf{Num\_MV} & \textbf{Attack\_Method} & \textbf{Training CRPs} & \textbf{Success Rate} \\ \hline
\multirow{10}{*}{5}  & MOO\_10     & 1m & 60\% \\ \cline{2-4} 
                     & MOO\_20     & 1m & 80\% \\ \cline{2-4} 
                     & MOO\_50     & 1m & 90\% \\ \cline{2-4} 
                     & MOO\_100    & 1m & 90\% \\ \cline{2-4} 
                     & MOO\_1000   & 1m & 90\% \\ \cline{2-4} 
                     & ALSCA\_10   & 1m & 85\% \\ \cline{2-4} 
                     & ALSCA\_20   & 1m & 90\% \\ \cline{2-4} 
                     & ALSCA\_50   & 1m & 80\% \\ \cline{2-4} 
                     & ALSCA\_100  & 1m & 60\% \\ \cline{2-4} 
                     & ALSCA\_1000 & 1m & 10\% \\ \hline
\multirow{10}{*}{20} & MOO\_10     & 1m & 0\%  \\ \cline{2-4} 
                     & MOO\_20     & 1m & 0\%  \\ \cline{2-4} 
                     & MOO\_50     & 1m & 5\%  \\ \cline{2-4} 
                     & MOO\_100    & 1m & 5\%  \\ \cline{2-4} 
                     & MOO\_1000   & 1m & 5\%  \\ \cline{2-4} 
                     & ALSCA\_10   & 1m & 5\%  \\ \cline{2-4} 
                     & ALSCA\_20   & 1m & 10\% \\ \cline{2-4} 
                     & ALSCA\_50   & 1m & 10\% \\ \cline{2-4} 
                     & ALSCA\_100  & 1m & 10\% \\ \cline{2-4} 
                     & ALSCA\_1000 & 1m & 0\%  \\ \hline
\multirow{10}{*}{50} & MOO\_10     & 1m & 0\%  \\ \cline{2-4} 
                     & MOO\_20     & 1m & 0\%  \\ \cline{2-4} 
                     & MOO\_50     & 1m & 0\%  \\ \cline{2-4} 
                     & MOO\_100    & 1m & 0\%  \\ \cline{2-4} 
                     & MOO\_1000   & 1m & 0\%  \\ \cline{2-4} 
                     & ALSCA\_10   & 1m & 0\%  \\ \cline{2-4} 
                     & ALSCA\_20   & 1m & 0\%  \\ \cline{2-4} 
                     & ALSCA\_50   & 1m & 0\%  \\ \cline{2-4} 
                     & ALSCA\_100  & 1m & 0\%  \\ \cline{2-4} 
                     & ALSCA\_1000 & 1m & 0\%  \\ \hline
\end{tabular}
\end{table}

\begin{table}
\centering
\caption{Attacking majority-voting-enhanced 64-bit (1,8)-IPUFs \\
using the ALScA method\cite{9973338} and the MOO method \cite{tobisch2020combining}}
\label{puf2}
\linespread{1.3}\selectfont
\setlength\tabcolsep{5pt}
\begin{tabular}{|c|c|c|c|}
\hline
\textbf{Num\_MV} & \textbf{Attack\_Method} & \textbf{Training CRPs} & \textbf{Success Rate} \\ \hline
\multirow{10}{*}{5}  & MOO\_10     & 1m & 10\% \\ \cline{2-4} 
                     & MOO\_20     & 1m & 30\% \\ \cline{2-4} 
                     & MOO\_50     & 1m & 50\% \\ \cline{2-4} 
                     & MOO\_100    & 1m & 50\% \\ \cline{2-4} 
                     & MOO\_1000   & 1m & 50\% \\ \cline{2-4} 
                     & ALSCA\_10   & 1m & 70\% \\ \cline{2-4} 
                     & ALSCA\_20   & 1m & 85\% \\ \cline{2-4} 
                     & ALSCA\_50   & 1m & 80\% \\ \cline{2-4} 
                     & ALSCA\_100  & 1m & 40\% \\ \cline{2-4} 
                     & ALSCA\_1000 & 1m & 20\% \\ \hline
\multirow{10}{*}{20} & MOO\_10     & 1m & 0\%  \\ \cline{2-4} 
                     & MOO\_20     & 1m & 0\%  \\ \cline{2-4} 
                     & MOO\_50     & 1m & 5\%  \\ \cline{2-4} 
                     & MOO\_100    & 1m & 5\%  \\ \cline{2-4} 
                     & MOO\_1000   & 1m & 5\%  \\ \cline{2-4} 
                     & ALSCA\_10   & 1m & 10\% \\ \cline{2-4} 
                     & ALSCA\_20   & 1m & 20\% \\ \cline{2-4} 
                     & ALSCA\_50   & 1m & 20\% \\ \cline{2-4} 
                     & ALSCA\_100  & 1m & 10\% \\ \cline{2-4} 
                     & ALSCA\_1000 & 1m & 0\%  \\ \hline
\multirow{10}{*}{50} & MOO\_10     & 1m & 0\%  \\ \cline{2-4} 
                     & MOO\_20     & 1m & 0\%  \\ \cline{2-4} 
                     & MOO\_50     & 1m & 0\%  \\ \cline{2-4} 
                     & MOO\_100    & 1m & 0\%  \\ \cline{2-4} 
                     & MOO\_1000   & 1m & 0\%  \\ \cline{2-4} 
                     & ALSCA\_10   & 1m & 0\%  \\ \cline{2-4} 
                     & ALSCA\_20   & 1m & 0\%  \\ \cline{2-4} 
                     & ALSCA\_50   & 1m & 0\%  \\ \cline{2-4} 
                     & ALSCA\_100  & 1m & 0\%  \\ \cline{2-4} 
                     & ALSCA\_1000 & 1m & 0\%  \\ \hline
\end{tabular}

\end{table}

    \textbf{Experiment Setup:} Our analysis centered on a series of hardware-generated 64-bit 10-XOR-PUFs and (1,8)-iPUFs, all subjected to majority voting to enhance their reliability. We employed current reliability-based attack methodologies for this analysis, notably the ALSCA method\cite{9973338} and the Multi-objectives-optimization method of Tobisch et al.\cite{tobisch2020combining}. In our experiments, 'Num\_MV' denotes the number of iterations of majority voting applied to each PUF. Higher numbers of majority voting iterations correlate with increased PUF reliability. It's important to note that in the accompanying tables, 'Attack\_Method' refers to the method being used and the number of times each Challenge-Response Pair is repeated during the attack process. This repetition is essential for gathering sufficient data to assess the PUF's reliability accurately. For example, "MOO\_10" denotes the Multi-objectives-optimization method with each training CRP being repeated 10 times. The detailed experimental setup, including the generation of CRPs and the specifics of the modeling, will be elaborated in a later section dedicated to experimental methodology.
    
   \textbf{Results and Analysis:} Experimental results are shown in Table~\ref{puf1} and Table~\ref{puf2}. The ALScA method, 
   has exhibited effectiveness for lower repeats of majority voting (5 times). However, its effectiveness dwindled with increased repeats of majority voting. Notably, as the repetition of CRPs increased, there was a discernible decline in performance. In parallel, the Multi-objectives-optimization method, leveraging a net difference approach, demonstrated capability in scenarios with standard reliability but struggled in high-reliability contexts. The increase of CRP repetition times did not significantly impact its performance. These observations underscore that both methods do not achieve enhanced reliability assessment accuracy by merely increasing CRP repetition times.

\section{A Low-Dimension High-Fidelity Reliability Representation}\label{sec_LDHF}

\subsection{Reasons for Ineffectiveness of Existing Neural Networks for Attacking Majority-Voting-Enhanced PUFs}\label{sec5.1}

The experimental attack results presented in Sec.~\ref{sec_defenseTestData} show the effectiveness of majority voting (MV) as a defense tool against the two most powerful reliability-based ML attack methods. Since ALScA is easier to use and more broadly applicable, in this section we are interested in finding out the main reasons behind the success of MV in fighting the ALScA method. 

While it can be easily analyzed that MV does not eliminate unreliability, we wonder if the success of MV reported in Sec.~\ref{sec_defenseTestData} is due to the inability of those attack methods to capture unreliability information, in the sense that all responses to a challenge stay the same within the number of repeated response measurements for a challenge. 

\subsubsection{Examining Attacks' Ability to Capture Unreliability Information}

To answer this question, we choose a set of $N$ randomly selected challenges (with $N$\,=\,1 million) and apply each challenge $m$ times (with $m$ going through 20, 100, 1000) to 50-repeat majority-voting-enhanced PUFs to get $m$ responses for each challenge from each PUF instance. For each challenge, the frequency of its responses being 1 is calculated and denoted by $f_1$, and the experimentally measured reliability of the PUF for the challenge $c$ is calculated by
\begin{equation} \label{reliability_challenge}
R_{m}(c) = |2f_1 - 1|,
\end{equation}
where $|\cdot|$ denotes the absolute value. We wish to comment that experimentally measured reliability (\ref{reliability_challenge}) for a challenge can also be defined using $f_0$, the frequency of responses being 0, by replacing $f_1$ in (\ref{reliability_challenge}) by $f_0$, and the two definitions are equivalent.
With the definition (\ref{reliability_challenge}), we then calculate the experimentally measured reliability of the PUF by
\[
R_{m}(PUF) =  \frac{\sum_{i=1}^{N} R_{m}(c_i)}{N},
\]
where $c_i$ is the $i$-th challenge. We listed the average experiment reliability of all tested PUF instances for the indicated PUF designs in Table~\ref{tab_ExperimentReliab}. 

\begin{table}[htbp]
\centering
\caption{Measured Reliability of PUFs Enhanced with 50-repeat-MV}
\label{tab_ExperimentReliab}
\linespread{1.3}\selectfont
\setlength\tabcolsep{5pt}
\begin{tabular}{|c|c|c|}
\hline
\textbf{PUF Type}           &\textbf{$m$} & \textbf{$R_{m}(PUF)$}  \\ \hline
\multirow{3}{*}{(1,8)-IPUF} & 20          & 0.931                  \\ \cline{2-3}
                            & 100         & 0.969                  \\ \cline{2-3}
                            & 1000        & 0.978                  \\ \hline
\multirow{3}{*}{(6,6)-IPUF} & 20          & 0.931                  \\ \cline{2-3}
                            & 100         & 0.970                  \\ \cline{2-3}
                            & 1000        & 0.979                  \\ \hline
\multirow{3}{*}{10-XPUF}    & 20          & 0.927                  \\ \cline{2-3}
                            & 100         & 0.964                  \\ \cline{2-3}
                            & 1000        & 0.974                  \\ \hline
%
\end{tabular}
\end{table}

The data listed in Table~\ref{tab_ExperimentReliab} show that unreliability was captured, to different extents, by the ALScA attack methods for $m=20, 100, 1000$, which have answered the question we raised in the second paragraph in Sec.~\ref{sec5.1}.

While unreliability information has been picked up by the representation of the ALScA as data in Table~\ref{tab_ExperimentReliab} show, we are still wondering if the captured unreliability is not accurate enough, a reason we suspect for the failure of attack methods for 50-MV-enhanced PUFs. An immediate challenge for this question is how to find the ground truth of unreliability information, or at least highly accurate unreliability information as an approximate ground truth, so that we can compare captured unreliability against the approximate ground truth.

It is generally accepted that when $m$ increases, the experimentally measured reliability converges to the true reliability, and $R_{m}$ with larger $m$ should more accurately reflect the true reliability. Thus, we consider $R_{m}$ with $m=1000$ as a high-accuracy approximation of the true reliability and calculated the mean difference of reliability (\ref{reliability_challenge}) at $m=20$ with the reliability at $m=1000$ by
$\sum_{i=1}^{N} |R_{20}(c_i)-R_{1000}(c_i)|/N$.
We also calculated the mean difference of reliability at $m=100$ with the reliability at $m=1000$, and the two mean differences for different PUFs and different MV repetitions are listed in Table~\ref{tab_ExperimentReliab2}. Data in the table do show that the ALScA with a lower $m$ value has a larger error in capturing unreliability information, but the error is below 6\% for $m=20$ and below 1.5\% for $m=100$. Thus, we believe the reliability representation of the ALScA can capture most of the unreliability information.

\begin{table}
\centering
\caption{Mean Difference of Measured Reliability at $m=1000$ with \\
         Measured Reliability at Different $m$ values}
\label{tab_ExperimentReliab2}
\linespread{1.3}\selectfont
\setlength\tabcolsep{5pt}
\begin{tabular}{|c|c|c|}
\hline
\textbf{PUF Type}           & \textbf{CRP\_Repetition} & \textbf{Mean Difference} \\ \hline
 \multirow{2}{*}{(1,8)-IPUF}& 20               & 0.054       \\ \cline{2-3}
                            & 100              & 0.011       \\ \hline
 \multirow{2}{*}{(6,6)-IPUF}& 20               & 0.052       \\ \cline{2-3}
                            & 100              & 0.011       \\ \hline
 \multirow{2}{*}{10-XPUF}   & 20               & 0.053       \\ \cline{2-3}
                            & 100              & 0.012       \\ \hline
\end{tabular}
\end{table}


\subsubsection{The Factor for Neural Networks Failure against MV}

Results in Sec.\,\,\ref{sec4_defense} show that the ALScA neural network attack method with $m=20, 100, 1000$ all failed for MV with 50 repeats. We suspect that attack methods with $m=20$ could possibly be incapable of capturing accurate unreliability information, but for $m=100, 1000$, there must be other reasons behind the success of majority-voting against those attacks. A possible reason for MV's success, we believe, is the diminishing learning power of neural network methods when the number of output nodes becomes large even if the true unreliability is accurately captured. 

Out belief on the large number of output nodes as a possible reason is based on the observation that 
the MLMSA and the ALScA methods, when each challenge is fed $m$ times to the PUF, use one-hot encoding to represent the experimentally measured reliability by an ($m\!+\!1$)-bit vector with the $i$-th bit being 1 if and only if the number of 1's is $i$ for $i=0, 1, \cdots, m$, which will lead to a large number of output nodes when $m$ is large.
It is highly likely that with $m=1000$ the experimentally measured reliability accurately approximates the true reliability, but the output of the MLMSA and the ALScA has 1002 output nodes with 1001 nodes representing the reliability information. 

To examine the relation between the number of output nodes and the learning power of neural network, we use the same neural network architecture of the ALScA method with $m=1000$, which was used for the attacks presented in Sec.~\ref{sec_defenseTestData}, but use the following information-losing $(k\!+\!1)$-dimensional representation of reliability: 

\begin{itemize}
    \item[] {\em For reliability information obtained from applying each challenge $m$ times to the PUF, we represent the experimentally measured reliability by a $(k\!+\!1)$-bit vector with one-hot encoding (assuming that $m$ is divisible by), and the $i$-th bit is 1 if and only if the count of 1-valued responses 
    is in the set of integers
\begin{equation}\label{lossy_representation}
\begin{aligned}
\hspace*{-3mm}
\left\{\!\frac{m\,i}{k}, \frac{m\,i\!+\!k}{k},\cdots, \frac{m\,i\!+\!m\!-\!k}{k}\right\}\mbox{ for }&i=0,\cdots\!, k\!-\!1,\\
\hspace*{-3mm}
\left\{m \right\}~~~~~~~~~~~~~~~~~~\mbox{ for }& i=k
\end{aligned}
\end{equation}
}
\end{itemize}

\begin{figure}[h]
\centering
   \includegraphics[width=9cm]{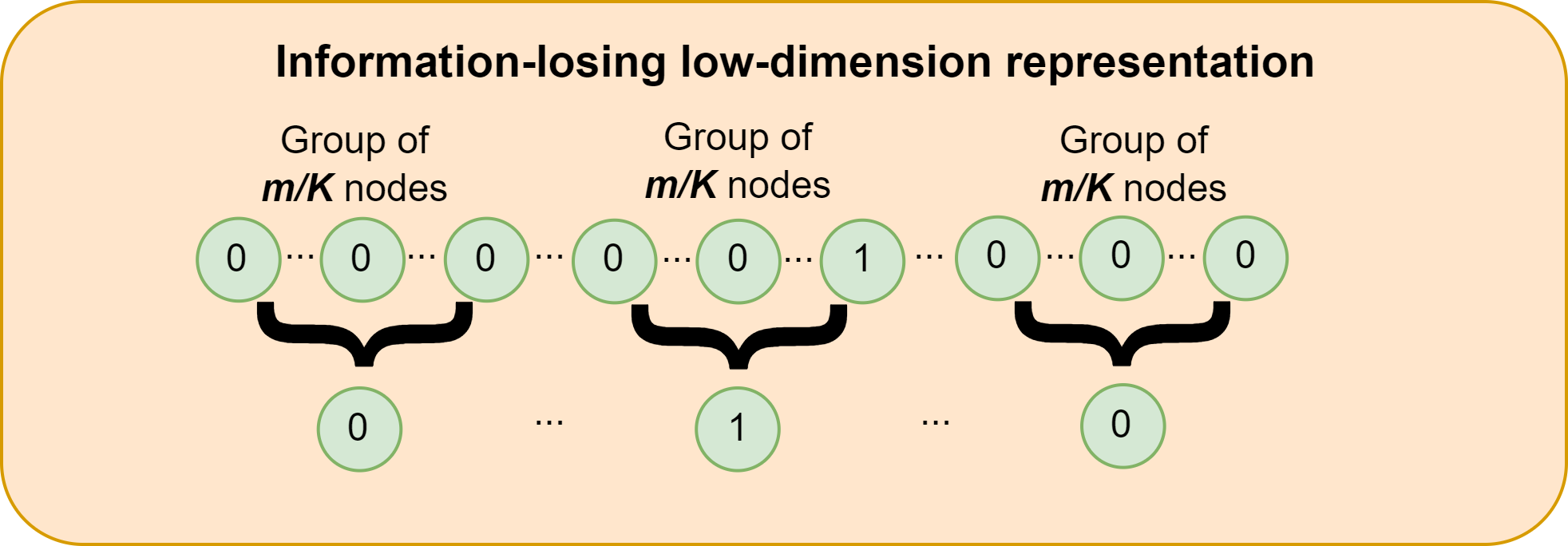}
\caption{Illustration of the information-losing low-dimension representation}
\label{HARR}
\end{figure}

This representation, for most cases, groups $m/k$ outcomes under the ALScA representation into one outcome (as illustrated in Fig.\,\,\ref{HARR}), and hence captures less unreliability information than the one used by the ALScA. 

To assess the power of attack methods using this representation, we used two information-losing representations of the type (\ref{lossy_representation}) with $(m,k)\!=\!(1000,10)$ and $(1000,20)$ respectively to replace the original reliability representation of ALScA. The two newly obtained methods were used to attack the same PUFs reported in Sec.~\ref{sec_defenseTestData}, and the attack results are listed in Table~\ref{lossy_representation_attacks}.

\begin{table}[htbp]
\centering
\caption{Attacks on 64-bit MV-enhanced 10-XOR-PUFs and (1,8)-IPUFs using the information-losing representation of reliability}
\label{lossy_representation_attacks}
\linespread{1.3}\selectfont
\setlength\tabcolsep{5pt}
\begin{tabular}{|c|c|c|c|c|}
\hline
\textbf{PUF type}                    & \textbf{Num\_MV}    & \textbf{k} & \textbf{CRPs} & \textbf{Success Rate} \\ \hline
\multirow{6}{*}{\textbf{10-XPUF}}    & \multirow{2}{*}{5}  & 10         & 1m            & 90\%                  \\ \cline{3-5} 
                                     &                     & 20         & 1m            & 90\%                  \\ \cline{2-5} 
                                     & \multirow{2}{*}{20} & 10         & 1m            & 30\%                  \\ \cline{3-5} 
                                     &                     & 20         & 1m            & 40\%                  \\ \cline{2-5} 
                                     & \multirow{2}{*}{50} & 10         & 1m            & 5\%                   \\ \cline{3-5} 
                                     &                     & 20         & 1m            & 5\%                   \\ \hline 
\multirow{6}{*}{\textbf{(1,8)-iPUF}} & \multirow{2}{*}{5}  & 10         & 1m            & 80\%                  \\ \cline{3-5} 
                                     &                     & 20         & 1m            & 80\%                  \\ \cline{2-5} 
                                     & \multirow{2}{*}{20} & 10         & 1m            & 40\%                  \\ \cline{3-5} 
                                     &                     & 20         & 1m            & 40\%                  \\ \cline{2-5} 
                                     & \multirow{2}{*}{50} & 10         & 1m            & 0\%                   \\ \cline{3-5} 
                                     &                     & 20         & 1m            & 0\%                   \\ \hline 
\end{tabular}
\end{table}

The data listed in Table~\ref{lossy_representation_attacks} show that when $k=10$ or $20$, for all MV repeat values the modeling powers of the method with either of the two new representations are higher than that of the original ALScA\_1000 method in Tables~\ref{puf1} and \ref{puf2}, where ALScA\_$m$ denotes the ALScA method with each challenge applied $m$ times.
In addition, as data in Tables~\ref{puf1} and \ref{puf2} show, the modeling powers of the new methods are also much higher than the best of the MOO and ALScA methods for the 20-repeat MV, and has similar modeling power as the best of the MOO and ALScA methods for the 5-repeat MV.  Each of the two new representations obviously captures less unreliability information than that of the ALScA\_1000, and then it is highly likely that their higher learning powers come from the {\em lower output dimensionality} when compared with that of the ALScA. 

\subsection{The Proposed Reliability Representation} 

A reliability-based machine learning attack method usually takes a challenge as input and uses the PUF reliability for the challenge, in some representation, as part of the output. For reliability-based attacks where each challenge is repeatedly applied $m$ times, one of the representations of the lowest dimensionality is the experimentally measured reliability (\ref{reliability_challenge}) for the challenge, a one-dimensional representation.

But our goal is not to minimize the dimensionality of the representation, but to maximize, or approximately maximize, the learning power of the attack method. 
Among all reliability-based ML attack methods surveyed in Sec.\ref{sec3}, our comparative study in Sec.~\ref{sec3.C} shows that neural network methods possess broader applicability and higher modeling power than PUF-tailored methods. 
The MLMSA and the ALScA methods, by using different groups of nodes to represent different types of information, leverage techniques of multi-task learning to reach high and robust modeling power, and the ALScA method, with its neural network architecture adapted for reliability-based attacks, is the most effective machine learner among all methods surveyed in Sec.~\ref{sec3}.

While more information used by a learning method in general leads to higher learning performance, parameter values of a learning method, and in this case the dimensionality of the neural network output, also plays a significant role, as revealed by our study in Sec.~\ref{sec4_defense} on ALScA's failure in learning majority-voting-enhanced PUFs as well as by the study in Sec.~\ref{sec5.1} on ALScA's improved learning power when a lower-dimensional reliability representation is used, a representation which actually retains less reliability information than the original representation of the ALScA. The studies in Sec.~\ref{sec4_defense} and Sec.~\ref{sec5.1} motivate us to look for attack methods that may reach a new level of learning power if we can develop an information-preserving representation with a lower dimensionality, a dimensionality adjusted for maximizing learning power of the attack method. Our motivation is built upon the following observations:

\begin{itemize}
\item[1.] For reliability-based attacks where each challenge is repeatedly applied $m$ times, a representation of the highest dimensionalilty among all methods we surveyed in Sec.\,\,\ref{sec3} is the one used by he MLMSA and the ALScA methods, which use an ($m\!+\!1$)-dimensional vector of one-hot encoding to represent ($m\!+\!1$) possibilities for the count of 1-valued responses. 

\item[2.] A repesentation of the lowest-dimensionality 
is $f_1$, the frequency of 1's, which is implementable using one neural network node and can represent ($m\!+\!1$) possibilities with ($m\!+\!1$) different values from the set $\{ j/m: j=0, 1, \cdots, m\}$. Thus, by allowing a node to assume ($m\!+\!1$) possible values, $f_1$ retains the same amount of information carried by the ($m\!+\!1$)-dimensional representation of the MLMSA and the ALScA. 

\item[3.] Another representation also of the lowest dimensionalilty is the experimentally measured reliability (\ref{reliability_challenge}) for a challenge, but this repesentation retains only ``half'' information represented by $f_1$, since two values of $f_1$, e.g. $j/m$ and $(m\!-\!j)/m$, are mapped to one value of the reliability (\ref{reliability_challenge}).

\item[4.] The representation of ALScA and $f_1$ are the two extremes in terms of diemensionality. But other information-preserving representations of intermediate diemensionality could be of value if we are to search for a representation 
of a particular dimensionality to maximize the modeling power of attacks.
\end{itemize}

Along the line of thought expressed above, our strategy is to develop a set of representaions of experimentally measured reliability, with the dimensionality being the indexing variable for representations in this set, and this dimensionality is to be used as the very optimization variable in a procedure in which we look for a neural network attack method with optimal, or near optimal, learning performance. 

More specifically, for an attack which applies each challenge $m$ times to the attacked PUF, we are looking for a $(k\!+\!1)$-dim representation which retains the same, or similar, amount of information as $f_1$, with $k$ being a variable. Presented in the following are our technical thoughts which lead to the representation.

\begin{itemize}
\item When $m$ is a multiple of $k$, applying a challenge $m$ times to the PUF can be viewed as carrying out $m/k$ events, where each event consists of applying the challenge $k$ times to the PUF. Each event will produce a count of 1-valued responses out of $k$ responses, with the count ranging from 0 to $k$. 

\item If $m$ approaches infinity, a $(k\!+\!1)$-dim representation we have thought of is the $(k\!+\!1)$-dim vector
\begin{equation}\label{KD_IdealRepresent}
(p_0, p_1, p_2, \cdots, p_k),
\end{equation}
where $p_i$ is the probability that the count of 1-valued responses in an event is equal to $i$. 

\item When $m$ is far away from infinity and divisible by $k$, the following $(k\!+\!1)$-dim vector is a representation that approximates the one above:
\begin{equation}\label{KD_PracticalRepresent}
(b_0, b_1, b_2, \cdots, b_k),
\end{equation}
where $b_i$ is the frequency that the count of 1-valued responses is equal to $i$ in the $m/k$ events. 


\item When $m$ is not divisible by $k$, we choose the largest integer $m'$ that is less than $m$ but divisible by $k$, and the discussion in the preceding paragraph stays valid if $m$ is replaced by $m'$. This validity allows people to examine different $k$ values for an experiment study where each challenge is applied $m$ times without have to adjust the $m$ value for different $k$ value.
\end{itemize}

Summarizing the discussions above, we are ready to present our presentation for experimentally measured reliability of the PUF for a challenge $c$.

\vspace{0.5mm}
\begin{itemize}
\item[] {\bf\em The $(k\!+\!1)$-dim Representation of Measured Reliability}\vspace{0.5mm}
\item[] {\em For an experiment in which every randomly chosen challenge is applied $m$ times to the PUF, and for a positive integer $k$ smaller than $m$, let $m'$ denote the largest integer that is bounded by $m$ and is divisible by $k$, our choice to represent the experimentally measured reliability of the PUF for a challenge $c$ is the representation (\ref{KD_PracticalRepresent}) with $b_i$ being the frequency that the count of 1-valued responses is equal to $i$ in the $m'/k$ events.}\vspace{0.5mm}
\end{itemize}

It is clear that when $m$ is divisible by $k$, i.e. $m'=m$, our representation can reproduce all information representable by that of the ALScA\_$m$. But when $m$ is not divisible by $k$, our representation retains less information than that of the ALScA\_$m$, but carries the same amount of information representable by that of the ALScA\_$m'$. Since this representation allow people to use a lower-dimensional representation but retains the same or almost the same amount of information, we call measured reliability expressed in this representation the low-dimension high-fidelity (LDHF) reliability. 

Given in Algorithm~\ref{alg:reliability-assessment} is the procedure for calculating the LDHF reliability through an experiment of applying a challenge $m$ times to a PUF. Note that when $m$ is not divisible by $k$, the procedure will stop at the largest integer that is less than $m$ and divisible by $k$.
\begin{algorithm}
\caption{Procedure for Calculating LDHF Reliability}
\label{alg:reliability-assessment}
\begin{algorithmic}[1]
\State \textbf{Input:} Challenge $c$\,; Parameters $m$ and $k$\,.
\State \textbf{Output:} LDHF$[k\!+\!1]$: an array of size $(k\!+\!1)$ 
\Procedure{LDHF\_Reliability}{$c$, $m$, $k$}
\State Initialize the array LDHF to $(k\!+\!1)$ zeros

\State $iterations=0$
\For{($i=0; ~i<m;$~ increment $i$ by $k$)}
    \State Get $k$ responses from PUF with $c$ applied $k$ times
    \State $Count \gets$ number of 1-valued responses 
    \State If\,($Count$ is $j$) Increment LDHF$[j]$ by $1$
    \State Increment $iterations$ by $1$
\EndFor

\State Divide all elements of LDHF by $iterations$
\State \Return the array LDHF
\EndProcedure
\end{algorithmic}
\end{algorithm}


\section{Experimental Assessment of the LDHF Representation}\label{sec_experiment}

\subsection{Experimental Setup}
We carried out experimental attacks using neural network methods equipped with the proposed LDHF representation, and the experiments were set up as follows.

\begin{itemize}
    \item \textbf{PUF Implementation and Operating Conditions:} PUF instances were realized in VHDL with Xilinx Vivado, and challenges were processed and responses recorded via the Xilinx SDK. Standardized operating conditions were maintained for test consistency.
    
    \item \textbf{CRP Generation:} CRPs were produced using Xilinx Artix\textregistered-7 FPGAs, with each PUF type tested across three distinct FPGA boards, creating 20 unique instances to ensure a broad range of test cases.
    
    \item \textbf{Attack Implementation, Training and Testing:} The attack methods were implemented using Python 3.7, TensorFlow, and Keras on a high-performance computing cluster. Due to the unavailability of the source code of the ALScA method \cite{9973338}, we developed the code based on the description in \cite{9973338} with adaptation to using only response and reliability information. For each attack, the CRP data was partitioned with a 90-10 split for training and testing, setting aside 20\% of the training set for validation. An attack was deemed successful if it achieved a testing accuracy above 85\%. Source codes for these attack experiments will be made openly accessible after the paper is accepted for publication.
\end{itemize}

\subsection{The Experimental Study and Results}\label{sec_Exp.2}

To evaluate the effectiveness of the LDHF representation, we replace the reliability information representation of the ALScA by the LDHF and the resulting neural network is our attack method we will use for our experimental assessment study. The LDHF reliability has two parameters $m$ and $k$. To ensure that the experiments will capture unreliability information of highly reliable PUFs, we choose 1000 for $m$, that is, each challenge will be applied 1000 times to a PUF in a reliability attack. To determine a near-optimal value for $k$, the dimension of the represented reliability, we ran the attack method on 64-bit 10-XOR PUFs enhanced different levels of majority voting, and the attack results indicating $20$ is a reasonably good choice for the parameter $k$ for all tested cases.  We then decided to use the parameter set $(m,k)=(1000, 20)$ for all further experiments.

We applied the neural network attack method equipped with the LDHF reliability representation on XOR PUFs and IPUFs of different circuit architectural parameters enhanced different levels of majority voting. The attack results are listed in Table~\ref{LDHF_results}.

\begin{table}[htbp]
\centering
\caption{Attack study using neural networks equipped with \\the LDHF reliability on PUFs enhanced with \\different levels of majority voting}
\label{LDHF_results}
\linespread{1.3}\selectfont
\setlength\tabcolsep{5pt}
\begin{tabular}{|c|c|c|c|}
\hline
\textbf{PUF Type}                   &\textbf{Num\_MV}&\textbf{CRPs}& \textbf{Success Rate} \\ \hline
\multirow{3}{*}{\textbf{10-XPUF}}   &        5       &       1m    & 100\% \\ \cline{2-4} 
                                    &       20       &       1m    & 90\%  \\ \cline{2-4} 
                                    &       50       &       1m    & 90\%  \\ \hline
\multirow{3}{*}{\textbf{12-XPUF}}   &        5       &       1m    & 90\%  \\ \cline{2-4} 
                                    &       20       &       1m    & 90\%  \\ \cline{2-4} 
                                    &       50       &       1m    & 85\%  \\ \hline
\multirow{3}{*}{\textbf{(1,8)-IPUF}}&        5       &       1m    & 100\% \\ \cline{2-4} 
                                    &       20       &       1m    & 90\%  \\ \cline{2-4} 
                                    &       50       &       1m    & 85\%  \\ \hline
\multirow{3}{*}{\textbf{(6,6)-IPUF}}&        5       &       1m    & 90\%  \\ \cline{2-4} 
                                    &       20       &       1m    & 90\%  \\ \cline{2-4} 
                                    &       50       &       1m    & 80\%  \\ \hline
\end{tabular}
\end{table}

Comparing data in Table~\ref{LDHF_results} and data in Table~\ref{lossy_representation_attacks}, we can see that neural networks with the LDHF representation have substantially higher modeling power than networks with the information-losing representation presented in Sec.~\ref{sec5.1} for the 10-XOR PUF and the (1,8)-IPUF for all levels of majority voting. Please note that attack methods with the information-losing representation with $k=10$ or $20$ were already showing similar or higher modeling power than the best of the ALScA and the MOO method as shown by data in Table~\ref{puf1} and Table~\ref{puf2}. For the 12-XOR PUF and (6,6)-IPUF on which we did not launch attacks using existing methods or methods with the information losing representation, data in Table~\ref{LDHF_results} also show the LDHF representation is highly effective. 

Since majority-voting has been shown in Sec.~\ref{sec4_defense} to provide effective defense against existing reliability-based attacks, we are curious if the new attack method equipped with the LDHF reliability can be defeated by high-repeat majority-voting. To this end, we generated CRPs from MV-enhanced PUFs with 100-repeat MV, 200-repeat MV, or 500-repeat MV, and apply our attack method to the PUFs. The attack results are listed in Table~\ref{LDHFresults2}, showing that high-repeat majority voting cannot adequately defend against the new attack method, though has led to show lower success rates for the attacks when compared with majority voting with 5, 20, or 50 repeats listed in Table~\ref{LDHF_results}.


\begin{table}[htbp]
\centering 
\caption{Attacking PUFs with high-levels of majority-voting using neural networks equipped with the LDHF reliability}
\label{LDHFresults2}
\linespread{1.3}\selectfont
\setlength\tabcolsep{5pt}
\begin{tabular}{|c|c|c|c|} \hline
\textbf{PUF type} & \textbf{Num\_MV} & \textbf{CRPs} & \textbf{Success Rate} \\ \hline
\multirow{3}{*}{\textbf{10-XPUF}}    & 100 & 1m & 60\% \\ \cline{2-4} 
                                     & 200 & 1m & 40\% \\ \cline{2-4} 
                                     & 500 & 1m & 40\% \\ \hline
\multirow{3}{*}{\textbf{(1,8)-IPUF}} & 100 & 1m & 50\% \\ \cline{2-4} 
                                     & 200 & 1m & 30\% \\ \cline{2-4} 
                                     & 500 & 1m & 20\% \\ \hline
\end{tabular}
\end{table}

\begin{table}[htbp]
\centering
\caption{Attack study using neural networks equipped with \\the LDHF reliability on simulated PUFs enhanced with \\different levels of simulated noise}
\label{LDHF_results_simulation}
\linespread{1.3}\selectfont
\setlength\tabcolsep{5pt}
\begin{tabular}{|c|c|c|c|c|}
\hline
\textbf{PUF Type}                   &\textbf{Noise level}&\textbf{BER}&\textbf{CRPs}& \textbf{Success Rate} \\ \hline
\multirow{3}{*}{\textbf{10-XPUF}}   &        0.01   &0.012    &       1m    & 100\% \\ \cline{2-4} 
                                    &       0.005   &    &       1m    & 90\%  \\ \cline{2-4} 
                                    &       0.001  &     &       1m    & 90\%  \\ \hline
\multirow{3}{*}{\textbf{12-XPUF}}   &        5   &    &       1m    & 90\%  \\ \cline{2-4} 
                                    &       20   &    &       1m    & 90\%  \\ \cline{2-4} 
                                    &       50   &    &       1m    & 85\%  \\ \hline
\multirow{3}{*}{\textbf{(1,8)-IPUF}}&        5   &    &       1m    & 100\% \\ \cline{2-4} 
                                    &       20  &     &       1m    & 90\%  \\ \cline{2-4} 
                                    &       50   &    &       1m    & 85\%  \\ \hline
\multirow{3}{*}{\textbf{(6,6)-IPUF}}&        5   &    &       1m    & 90\%  \\ \cline{2-4} 
                                    &       20  &     &       1m    & 90\%  \\ \cline{2-4} 
                                    &       50  &     &       1m    & 80\%  \\ \hline
\end{tabular}
\end{table}
\section{Conclusion}\label{sec_conclusion}

Implementable with low hardware overhead and operable with low power, PUFs have the potential as hardware primitives for implementing lightweight security protocols for resource-constrained IoT devices. But before a PUF design can be adopted in security applications, it is important to identify its all security vulnerabilities. One of the threats to PUFs is reliability-based machine learning attacks, which leverage the unreliability of a PUF's responses to some challenges to help infer the behavior of the PUF. Since PUFs inherently have some level of unreliability, the risk of susceptibility to reliability-based attacks does exist, and hence it is useful to have tools that can help detect PUFs' vulnerability to reliability-based attacks, and this paper is right on developing such a tool for PUFs of high reliability. 

Based on experimental examinations, we have discovered that majority-voting with 50-plus repeats can defeat all existing reliability-based machine learning attacks, and our further examination of existing methods revealed that the high dimensionality of reliability information used by a group of generally applicable attack methods is an important reason for the success of high-repeat majority-voting against the group of attack methods. These findings led us to exploring lower dimensional representation of reliability, resulting in a low-dimension high-fidelity reliability representation that enables successful discovery of some PUFs' vulnerabilities that have escaped detection by existing reliability-based machine learning attacks.

\section*{Acknowledgment}

The research was supported in part by the National Science Foundation under grant No.\,\,2103563. Computing resources in the High Performance Computing Center (HPCC) of Texas Tech University were used for some of the work.

\bibliographystyle{IEEEtran}
\bibliography{cite.bib}
\end{document}